\documentstyle[aps,multicol,epsf]{revtex}
\tightenlines
\newcommand{\be}{\begin{equation}}
\newcommand{\ee}{\end{equation}}
\newcommand{\ber}{\begin{eqnarray}}
\newcommand{\eer}{\end{eqnarray}}
\newcommand{\de}{\end{equation*}}
\newcommand{\cer}{\begin{eqnarray*}}
\newcommand{\der}{\end{eqnarray*}}

\draft
\begin{document}
\title{Reconstruction of SU(1,1) States \cite{byline}}
\author{G. S. Agarwal\cite{other} and J. Banerji}
\address{Physical
Research Laboratory, Navrangpura, Ahmedabad 380 009, India}
\maketitle
\begin{abstract}
We show how group symmetries can be used to reconstruct quantum states.
In our scheme for SU(1,1) states, the input field passes through
a non-degenerate parametric amplifier and one measures the probability of
finding the output state with a certain number (usually zero) of photons
in each mode. The density matrix in the Fock basis is retrieved
from the measured data by least squares method after singular value decomposition of the design matrix.
 Several illustrative examples involving the reconstruction of
a pair coherent state, a Perelomov coherent state, and a coherent
superposition
of pair coherent states are considered. \end{abstract}
\begin{multicols}{2}
\section{Introduction} The problem of
the reconstruction of quantum states was first considered in the
fifties by Fano \cite {jb1} and Pauli \cite {jb2}. Since a
quantum system is completely described by its density matrix, the
task is essentially to reconstruct the density matrix of a system
from information obtained by a set of measurements performed on an
ensemble of identically prepared systems. To that end, the seminal
work of Vogel and Risken \cite {jb3} showed that for a single
mode {\it optical} field, the histograms of quadrature amplitude
distributions measured by {\it homodyne} detection, is just the
Radon transform (or {\it tomography}) of the corresponding Wigner
function. One can thus obtain the Wigner function by taking the
inverse Radon transform of the data. Finally, the density matrix
in the position representation is obtained from the Wigner
function by Fourier transformation. This is the basis of {\it
optical homodyne tomography} \cite {jb3,jb4,jb5,jb6}. The
technique was experimentally realized by Smithey et al \cite
{jb4} who obtained the Wigner function and the density matrix
of vacuum and quadrature-squeezed states of a mode of the
electromagnetic field by using balanced homodyne detection. Much
progress has been achieved in this field over the last few years
\cite {jb6}. It is now well known, for example, that
 one can determine the density
matrix directly from the measured quadrature distribution without
having to evaluate the Wigner function. Additionally, parallel
tomographic schemes such as symplectic tomography \cite {jb7},
and photon number tomography \cite {jb8} have been suggested
for the reconstruction of quantum states of the light field which
can even be multi-mode \cite {jb9}. Other quantum  systems for
which reconstruction procedures were proposed include
one-dimensional wave packets \cite {jb10}, harmonic and
anharmonic molecular vibrations \cite {jb11}, motional states
of atom beams \cite {jb12}, Bose-Einstein condensates \cite
{jb13}, cyclotron states of a trapped electron \cite {jb14},
atomic Rydberg wave functions \cite {jb15}, atoms in optical
lattices \cite {lattice}, systems with a finite-dimensional state
space (e.g., for spin) \cite {jb17} and states in cavity QED
\cite {gsa3044}. Experimental reconstructions were reported for
electronic angular momentum states of hydrogen \cite {jb19},
vibrational quantum states of a diatomic molecule \cite {jb20}
and motional state of a single trapped atom \cite{prl4281_96}.
Vasilyev et al \cite{vasil} have reported tomographic measurement
of joint photon statistics of the two-mode quantum state produced
in parametric amplification.

While extensive work has been done on states of a two-mode field,
there are very many physical situations where the state to be
reconstructed has certain group symmetry. Clearly one could
benefit considerably by the use of the group symmetry properties
in the reconstruction of the state. For example, in the process of
down conversion, the two photons are produced together. In this
case the difference in the photon number in the two modes is
conserved and the state has the symmetry property of the SU(1,1)
group. In a previous publication, one of us has discussed how the
underlying SU(2) symmetry of a state can be utilized very
efficiently for its reconstruction \cite{gsaspin}. In this paper,
we consider reconstruction of states whose symmetry group is
SU(1,1) \cite{jbgsa}.
 The plan of the paper is as follows. In
section 2, we present a group theoretic perspective of a general
reconstruction procedure for quantum states. In section 3, we apply our
method to reconstruct some
important SU(1,1) states.
 The paper ends with concluding
remarks in section 4.
\section{Using Group Symmetries for State Reconstruction}
Let us first recall the principles of photon number tomography.
Several workers have suggested a procedure whereby the initial
state of the radiation field described by the density matrix
$\rho^{(in)}$, is displaced by different amounts. \ber
\rho^{(in)}\rightarrow \rho^{(out)} & = & {\cal
D}^{\dagger}(\alpha) \rho^{(in)} {\cal D}(\alpha),\nonumber\\
{\cal D}(\alpha) & = & \exp(\alpha a^\dagger -\alpha^* a), \eer
One then measures the distribution of photons in the displaced
field. The photon count in the output field is used to reconstruct
the $s$-ordered distribution function of the input field. This
method was very successfully used to measure
the vibrational state of a trapped ion \cite{prl4281_96}. There is
a related suggestion in the context of cavity QED which yields the
characteristic function of the radiation field \cite{gsa3044}. In
both these situations one measures atomic populations with rather
high efficiency. Though a direct photon counting measurement
suffers from questions of poor efficiency of photodetectors, there
exist several proposals on how to go around the problem
\cite{detec}.

For the two-mode field with SU(2) symmetry, one can displace the
state using the corresponding unitary operator for the SU(2)
group. This has been shown to enable one to reconstruct the states
of spin systems, states of polarization etc \cite{gsaspin}. This
is also closely related to a proposal in the context of Bose
Einstein condensates \cite{jb13}. The displacement of the state
is physically realized (say) by using external fields  in the case
of two-level atoms or spins. In the case of radiation fields such
a displacement is realized by optical components like waveplates
\cite{mandel}.

We next consider the case when the underlying symmetry of the
state is of the SU(1,1) group. The generators of this group are
\be
K_+=a^\dagger b^\dagger,\quad K_-=a b,\quad K_z=(a^\dagger
a+b^\dagger b+1)/2, \ee where $a^\dagger a-b^\dagger b$ = constant
= $q$ (say). Without any loss of generality, one can assume that
$q\geq 0$. In that case, the vacuum state is given by the two-mode
Fock state $|q,0\rangle$ with the property $K_-|q,0\rangle=0$. The
displacement operator for this group is the well known squeezing
operator  parametrized by a complex quantity $z$:
\be
{\cal S}(z)=\exp (z a^\dagger b^\dagger - z^* a b). \ee Acting on
the $\vert 0,0\rangle$ state, it produces the squeezed vacuum
state \be\label{four}
 |z\rangle_0={\cal S}(z)|0,0\rangle. \ee It
should be noted that even though we are dealing with the two-mode
field, the underlying symmetry makes ${\cal S}(z)$ different from
the product ${\cal D}(\alpha){\cal D}(\beta)$ of the displacement
operators. We can now proceed in the spirit of earlier
constructions for the Heisenberg-Weyl and the SU(2) groups. We
consider the operator defined by
\be
\rho^{(out)}={\cal S}^\dagger (z)\rho^{(in)}{\cal S}(z). \ee and
the measurement of (say) $q$ photons in mode $a$ and \underline{no
} photons in mode $b$, i.e., the quantity \ber p^{(out)}(q,0) & =
& \langle q,0\vert \rho^{(out)}\vert q,0\rangle\nonumber\\ & = &
\langle q,0\vert{\cal S}^\dagger(z)\rho^{(in)} {\cal S}(z)\vert
q,0\rangle\nonumber\\ & = & _q\langle z\vert\rho^{(in)}\vert
z\rangle_q\equiv {\cal Q}(q,z)\eer where $\vert z\rangle_q$ is
defined in analogy to Eq (\ref{four}) with $\vert 0,0\rangle$
replaced by $\vert q,0\rangle$. We would now like to demonstrate
how measurements of ${\cal Q}(q,z)$ for a range of values of $z$
can be used to reconstruct the input state $\rho^{(in)}$.
 In
this case, as indicated in Fig. 1, $\rho^{(out)}$ can be obtained
from $\rho^{(in)}$ by passing the input state through a
nondegenerate parametric amplifier whose action is described by
the Hamiltonian $H=\lambda a^\dagger b^\dagger + h.c$ where
$\lambda$ is related to the non-linear susceptibility. The
operator ${\cal S}(z)$ is simply the evolution operator for this
Hamiltonian with $z=i\lambda t$.
\begin{center}
\begin{figure}[h]
\epsfxsize 7.0 cm \centerline{\epsfbox{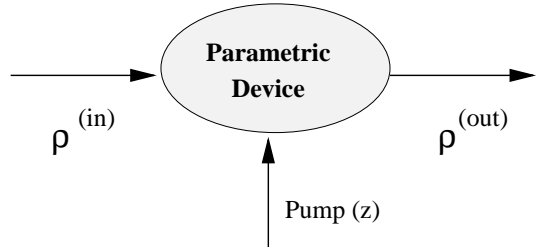}} \caption{
\narrowtext {Schematic of the reconstruction procedure.}}
\end{figure}
\end{center}
Using the disentangling theorem for ${\cal S}$ and substituting in
the expression for ${\cal Q}(q,z)$, we can write this probability
as a function of two auxiliary, {\it experimentally controlled}
parameters
\be
y=\tanh^2 |z|,\quad \phi=i\ln \left({z\over |z|}\right). \ee After
some algebra, one obtains \ber \lefteqn{ {\cal Q}(q,z)\equiv {\cal
Q}( q,y,\phi)={(1-y)^{q+1}\over q!}\times}\nonumber\\ & &
\sum_{m,n=0}^\infty \sqrt{{(m+q)! (n+q)!\over m!n!}}e^{i(m-n)\phi}
y^{(m+n)/2} \rho_{n,m}(q). \eer For the sake of clarity, we have
used the notation $\langle n+q,n\vert \rho^{(\rm in)}\vert
m+q,m\rangle=\rho_{n,m}(q)$. At this point, we make the physically
reasonable assumption that \be\label{nmax} \rho_{n,m}(q)\approx 0,
\quad\hbox{for}\quad m,n>n_{\rm max}, \ee if $n_{\rm max}$ is
suitably large \cite{vogel}. Next we introduce the Fourier
Transform of the probability data with respect to the phase angle
$\phi$: \be\label{ft} g_k(q,y) = \int_0^{2 \pi}{d\phi\over 2\pi}
e^{ik\phi} {\cal Q}(q,y,\phi), \ee and construct the quantity
\be
f_k(q,y)={g_k(q,y) y^{-k/2}\over (1-y)^{q+1}}. \ee The
construction is legitimate since the potentially singular points
$y=0$ and $y=1$ are inaccessible to the experimenter. The point
$y=0$ corresponds to `doing nothing' to the input state whereas
$y=1$ would correspond to $|z|$ (and hence, either the pump
amplitude or the duration of the experiment) being infinity.

The integration over $\phi$ yields a Kronecker delta function and
one obtains a simple power series expansion for $f_k(q,y)$:
\be
f_k(q,y)= \sum_{m=0}^{n_{\rm max}-k}B_{mk}(q) \rho_{m+k,m}(q)
y^m,\ee where \be B_{mk}(q)={1\over q!}\sqrt{{(m+k+q)! (m+q)!\over
m!(m+k)!}} . \ee The task now is to
 obtain the density matrix elements from
tabulated values of $f_k(q,y)$. This can be done, in principle, by
least squares inversion \cite{vogel}.
\subsubsection{Least Squares Method}
 We write $f_k (q,y)$ in the form $f_k(q,y)= \sum_{j=1}^M a_j^{(M)}
\phi_j(y)$ where  $\phi_j(y)=y^{j-1}$ are the basis functions, $
a_j^{(M)}= B_{j-1,k}(q)\rho_{j-1+k,j-1}(q)$ contain the unknown
density matrix elements, and $M=n_{\rm max} -k+1$. Here the
superscript $(M)$ denotes that the coefficients in general, depend
on the number of basis functions included in the approximation.
Let $\tilde{y}_1$, $\tilde{y}_2$,...,$\tilde{y}_N$ be a set of
points at which the values of $f_k (q,y)$ are measured. We denote
by $\tilde{f}_i$ the measured value at $\tilde{y}_i$ with an error
$\tilde{f}_i-f_k(q,\tilde{y}_i)$. It is generally assumed that the
error at different points is uncorrelated. The design matrix $G$
is an $N\times M$ matrix whose $ij$-th element is given by
$G_{ij}=\phi_j(\tilde{y}_i)$. We introduce two vectors
$\vec{a}=\{a_1, a_2, ....a_M\}$ and $\vec{b}=\{\tilde{f}_1,
\tilde{f}_2, ..... \tilde{f}_N\}$. In least squares method, the
coefficients $a_j$ are determined by minimizing the quantity
$\chi^2=\vert G\vec{a}-\vec{b}\vert^2$.

 Although the method of least squares finds extensive
use in literature, it will give meaningful values for the
coefficients $\rho_{m+k,m}(q)$ only for small values of $m$. This
is so because for large values of $m$, the corresponding normal
equations become ill-conditioned. Hence we cannot expect to solve
them unless very high precision arithmetic is used. Even then a
slight change in the data (due for example, to round-off error)
may change the solution significantly. This ill-conditioning can
be traced to the fact that for large values of $m$, the basis
functions $y^m$ are not really independent in the sense that there
will be little difference between terms of (say) $y^9$ and
$y^{10}$ if the precision in the measured data is unable to
resolve it. In such cases, one must use Tikhonov regularization or
Singular Value Decomposition (SVD) of the design matrix in order
to extract meaningful results for the unknown coefficients
$\rho_{m+k,m}(q)$. We adopt the SVD approach \cite{antia} in which
one works directly with the design matrix $G$ rather than with
$G^TG$ (as in the least squares method without SVD). Thus the
ill-conditioning gets much reduced. The design matrix $G$ is
written in the form $G=U\Sigma V^T$, where $U$ is a $N\times M$
matrix, $\Sigma$ is a $M\times M$ diagonal matrix with diagonal
elements $\sigma_1$, $\sigma_2$, .... $\sigma_M$, and $V$ is a
$M\times M$ orthogonal matrix so that $U^T U=V^TV=VV^T=I_M$, the
$M\times M$ unit matrix. The matrix $U$ consists of $M$
orthonormalised eigenvectors associated with the $M$ largest
eigenvalues of $GG^T$, and the matrix $V$ consists of the
orthonormalised eigenvectors of $G^TG$. The diagonal elements of
$\Sigma$ are the nonnegative square roots of the eigenvalues of
$G^TG$ and are called the {\it singular values}. If $\vec{u_i}$
and $\vec{v_i}$ are the $i$-th columns of $U$ and $V$
respectively, then the solution can be written as
$\vec{a}=\sum_{i=1}^M \left(\vec{u_i}\cdot\vec{
b}/\sigma_i\right)\vec{v_i}$. The variance in the estimated
parameters $a_j$ can be written as
$\sigma^2(a_j^{(M)})=\sum_{i=1}^Mv^2_{ji}/\sigma^2_i$. It can thus
be seen that the error will be rather large for small $\sigma_i$,
and dropping such terms will reduce the errors, at the cost of
increasing the mean square deviation slightly. The columns of $V$
 corresponding to small $\sigma_i$ identify the linear combination
of variables, which contribute little towards reducing $\chi^2$,
but make large contribution in the standard deviation. Thus even
if some of the singular values are not small enough to cause
round-off problems, it may be better to zero them while computing
the solution.
\section{Results and
Discussion} In this section we will reconstruct the density matrix
from a simulation of the corresponding probability data for a pair
coherent state \cite{gsapair}, a Perelomov \cite{perel} coherent
state and a coherent superposition of pair coherent states.

In a real experiment the parameters $y$ and $\phi$ can take only a
finite (however large) number of values. In the absence of any
{\it a priori} knowledge about the input state, we choose a set of
 values of $\phi$ equally distributed between $0$ and $2 \pi$:
$\phi_s=2\pi s/N_{\phi}$, and a set of values of $y$ which are
equi-spaced between $y_{\rm min}=0.1$ and $y_{\rm max}=0.9$:
$\tilde{y}_n=y_{\rm min}+(y_{\rm max}-y_{\rm min})(n-1)/(N-1)$. Then
the Fourier transform with respect to $\phi$ in Eq. (\ref{ft}) is
approximated by a discrete Fourier transform
\be
g_k(q,\tilde{y}_i)\rightarrow {1\over N_\phi}\sum_{s=0}^{N_\phi
-1} e^{2 \pi i k s/N_\phi} {\cal Q}(q,\tilde{y}_i,2\pi s/N_\phi).
\ee Thus apart from truncation error due to the assumption
(\ref{nmax}), one will also have to deal with error due to {\it
discretization} of the variables $y$ and $\phi$.  The systematic
error due to phase discretization can be reduced to zero by
choosing $N_\phi\geq 2 n_{\rm max}+1$ \cite{vogel} whereas the
error in the discretization of $y$ is of order $N^{-2}$ and can be
made arbitrarily small by taking a sufficiently large value of
$N$. We have set $N_\phi=20$ and $N=101$ in the calculations to
follow. The data was simulated in the following way. We add to the
exact probability data $f_k(q,\tilde{y}_i)$ an error term $\delta
 f_k(q,\tilde{y}_i)=R{\cal G}( f_k(q,\tilde{y}_i))\sqrt{ f_k(q,\tilde{y}_i)/\tau}$, where
$R$ is a real random
 number uniformly distributed between -1 and 1,
 ${\cal G}$ is a Gaussian distribution with
 zero mean and unit variance \cite{gaussian}, and  $\tau=20000$ is the
number of trials at
 $y=\tilde{y}_i$.
All our calculations have been carried out using the software
package {\it Mathematica}. For the record, the random numbers were
generated with a seed value of 45.

\subsection{Reconstruction of a pair coherent state}
Pair coherent states of the radiation field can be generated via
the competition of four-wave mixing and two-photon absorption in a
nonlinear medium \cite{gsapair}. Pair coherent states can also be
realized for the motion of a trapped ion \cite{knight}. One drives
the ion with a laser on resonance and two other lasers with
appropriately chosen directions of propagation and tuned to the
second lower vibrational side band. In the Lamb-Dicke limit, the
ion is found in a pair coherent state.

The state vector for a pair coherent state has the form
 \ber \vert\Phi(\xi,p)\rangle & = &
N(\xi,p)\sum_{n=0}^{\infty}{{\xi}^n \over\sqrt{n!(n+p)!}}\vert
n+p,n\rangle,\nonumber\\ N(\xi,p) & = &
\left[\sum_{n=0}^{\infty}{\vert{\xi}\vert^{2n}\over
n!(n+p)!}\right]^{-1/2}. \eer Here $\xi$ is a complex parameter
and $p\geq 0$ is an integer. The corresponding exact density
matrix elements in the Fock basis are given by
\be
\rho_{n,m}(p)=\vert N_p\vert^2 {{\xi}^n {{\xi}^*}^m\over
\sqrt{n!(n+p)! m! (m+p)!}} \ee Note that
$\rho_{m,n}(p)=\rho_{n,m}^*(p)$ and for real values of $\xi$, the
density matrix is symmetric.

The exact diagonal density matrix
elements for the state $\vert\Phi(3,0)\rangle$ are plotted in Fig.
2(a). The least squares reconstruction from the simulated data
fails in this case (some of the diagonal elements assume absolute
values of the order of $10^3$ or so!) even with SVD when the tolerance
parameter is set to its default value of $10^{-p+2}$ where $p$ is
the machine precision. The failure is due to {\it overfitting},
that is, the use of a higher degree polynomial for
$f_0(0,\tilde{y}_i)$ than necessary. As a result, the design
matrix becomes ill-conditioned and some of the diagonal elements
of $\Sigma$ become very small. We mention parenthetically that the
default tolerance removes none of these singular (or almost
singular) values.

The situation can be improved substantially by probing certain
diagnostic indicators. The sequential sum of squares, for example,
is useful in determining the degree of the univariate polynomial
model. Each element in the sequential sum of squares vector
corresponds to the increment in the model sum of squares obtained
by sequentially adding each (non-constant) basis function to the
model. It can be seen from Fig. 2(b) that the contribution of the
basis functions $y^m$ in this respect decreases rapidly and
monotonically up to $m=6$ and is negligible (albeit oscillatory)
thereafter, suggesting that the optimum degree of the polynomial
for $f_0(q,\tilde{y}_i)$ should be five or six.
\vspace*{-1cm}
\begin{center}
\begin{figure}[h]
\epsfxsize 7.0 cm \centerline{\epsfbox{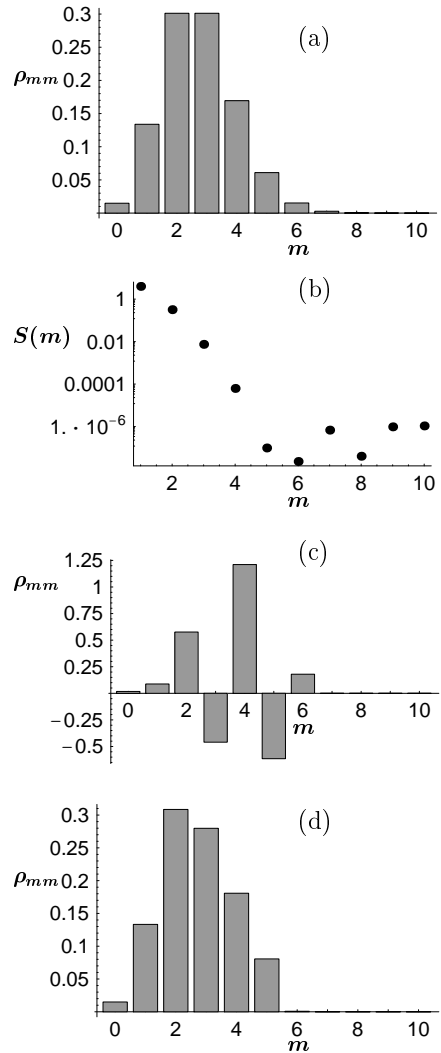}}
\caption{\narrowtext {Reconstruction of the diagonal density
matrix elements
 $ \rho_{mm}$ of the pair coherent state $\vert\Phi (3,0)\rangle$
 (see Eqs. (15) and (16))
 by least squares method. (a) Exact values; (b) sequential sum
 of squares $S(m)$ as a function of the order $m$ of the fitting
 polynomial suggests that the optimum degree of the polynomial
 should be five or six; (c) reconstruction by using a
 sixth degree fitting polynomial for $f_0(0,\tilde{y}_i)$; (d) as
 in (c), but with the singular values $<0.1$ removed.
}}
\end{figure}
\end{center}
\vspace*{-0.5cm} Furthermore, if the errors are uncorrelated, then
the residuals $r_i=\tilde{f}_i-\sum_{j=1}^M a_j^{(M)}
\phi_j(\tilde{y}_i)$ should be randomly distributed and if we
count the number of sign changes $S_c$ in the sequence {$r_1$,
$r_2$, ....$r_N$}, then we should find a value close to $N/2$
(that is, within about $\sqrt{N/2}$). Thus, after adding every
term, we can check the number of sign changes and decide to
terminate the process when this number increases to its limiting
value. In the fitting of $f_0(0,\tilde{y}_i)$, the values of $S_c$
are found to be  3, 36, 49, 49, 49, 54, 51, 52 and 55 when the
degree of the fitted polynomial is increased from 2 to 10 in steps
of unity. With 101 data points, it can be seen that $S_c$ reaches
its terminal value when the degree of the polynomial is $\geq 4$.

Based on these observations, we  fit $f_0(0,\tilde{y}_i)$ by a
6-th degree polynomial instead of the 10th degree polynomial
originally implied in Eq. (12). This does not mean that $n_{\rm
max}$ is re-set to six. It simply means that reliable estimates
cannot be made for $\rho_{m,m}$ if $m$ is greater than six. The
reconstructed diagonal density matrix elements as plotted in Fig.
2(c) are still not quite in agreement with the exact results but
are at least of the same order of magnitude. The diagonal elements
of the $7\times 7$ matrix $\Sigma$ have the values {$12.2716$,
$4.00265$, $0.964488$, $0.176915$, $0.0245331$, $0.00247435$,
$0.00015933$}. Setting the smallest three elements {\it and} their
inverses to zero, one obtains a more realistic reconstruction of
the diagonal density matrix elements (Fig. 2(d)).
\vspace*{-1cm}
\begin{center}
\begin{figure}[h]
\epsfxsize 7.0 cm \centerline{\epsfbox{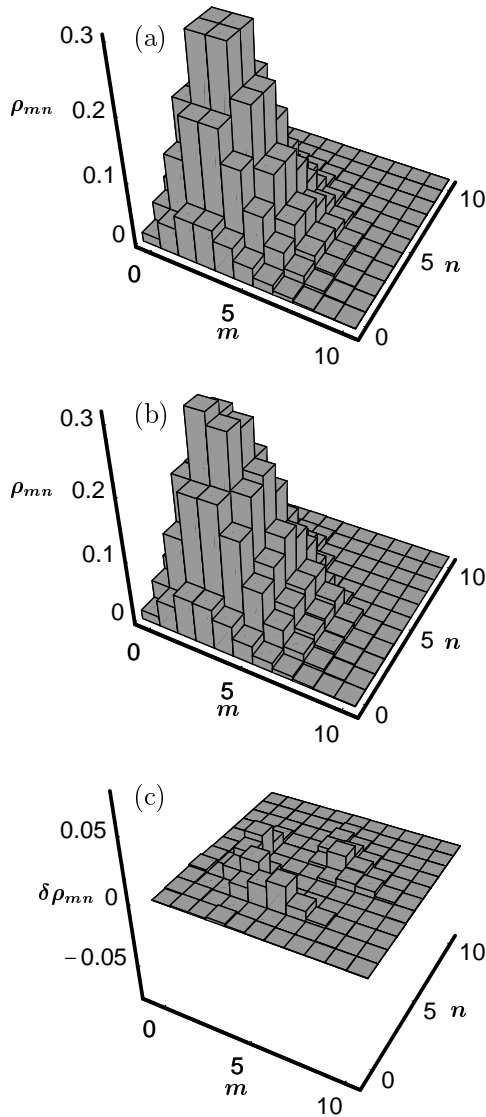}}
\caption{\narrowtext{Reconstruction of the density matrix elements
 $ \rho_{mn}$ of the pair coherent state $\vert\Phi (3,0)\rangle$
 by least squares method after singular value decomposition of
 the design matrix. The truncation parameter was set at $n_{\rm max}=10$.
 (a) Exact values; (b) reconstructed values;
 (c) the difference between the exact and the reconstructed
 values.}}
\end{figure}
\end{center}
\vspace*{-1cm}
We note that overfitting occurs only for values of
$k$ close to zero and is worst for the diagonal elements. A
systematic and consistent repetition of the above exercise for
other values of $k$ reveals that the optimum degree of the
polynomial is $5$ for $k=1$ and $2$, $4$ for $3\leq k\leq 5$ and
is $n_{\rm max}-k$ for $6\leq k\leq n_{\rm max}$. Removing the
singular values $<0.1$ in each case, we can finally reconstruct
all the elements of the density matrix. The result is in
reasonable agreement with the exact density matrix elements as
seen in Fig. 3.
\subsection{Reconstruction of a Perelomov coherent state}It is well known that
Perelomov coherent states can be produced in parametric
interactions. The state vector for a Perelomov coherent state is
given by
\be
\vert\Psi(\eta,q)\rangle={(1-\vert\eta\vert^2)^{{q+1\over 2}}\over
\sqrt{q!}}\sum_{p=0}^{\infty}\eta^p \sqrt{{(p+q)!\over p!}}\vert
p+q,p\rangle, \ee where $\eta$ is, in general, a complex parameter
with $|\eta|<1$, and $q\geq 0$ is an integer. The corresponding
exact density matrix elements in the Fock basis have the
expression
\be
\rho_{n,m}(q)={(1-\vert\eta\vert^2)^{q+1}\over q!}
\sqrt{{(n+q)!(m+q)!\over n!m!}}\eta^n{\eta^*}^m. \ee For $q=0$ and
real values of $\eta$, the density matrix is not only symmetric
but also has the following additional symmetries:
$\rho_{n+2k,n}(0)=\rho_{n+k,n+k}(0)$ and
$\rho_{n+2k+1,n}(0)=\rho_{n+k+1,n+k}(0)$. Consequently, only
$f_0(0, y)$ and $f_1(0,y)$ need to be measured and modeled. We
choose $q=0$, $\eta=0.6$ and set $n_{\rm max}=10$. Proceeding as
before, we plot the exact density matrix elements in Fig. 4 along
with the computed elements reconstructed by the least squares
method with singular value decomposition. Once again, the
reconstruction is found to be satisfactory. The error in the
computed elements is seen to be slightly less than in the case of
pair coherent states as only 4th order fitting polynomials were
necessary in the present case. Using the method of optical
homodyne tomography, Vasilyev et al \cite{vasil} have, for the
first time, reconstructed the diagonal elements of the two-mode
Perelomov coherent state produced by a parametric amplifier. Their
experiment also demonstrates how well the SU(1,1) symmetry holds
in parametric amplification.
\vspace*{-1cm}
\begin{center}
\begin{figure}[h]
\epsfxsize 7.0 cm \centerline{\epsfbox{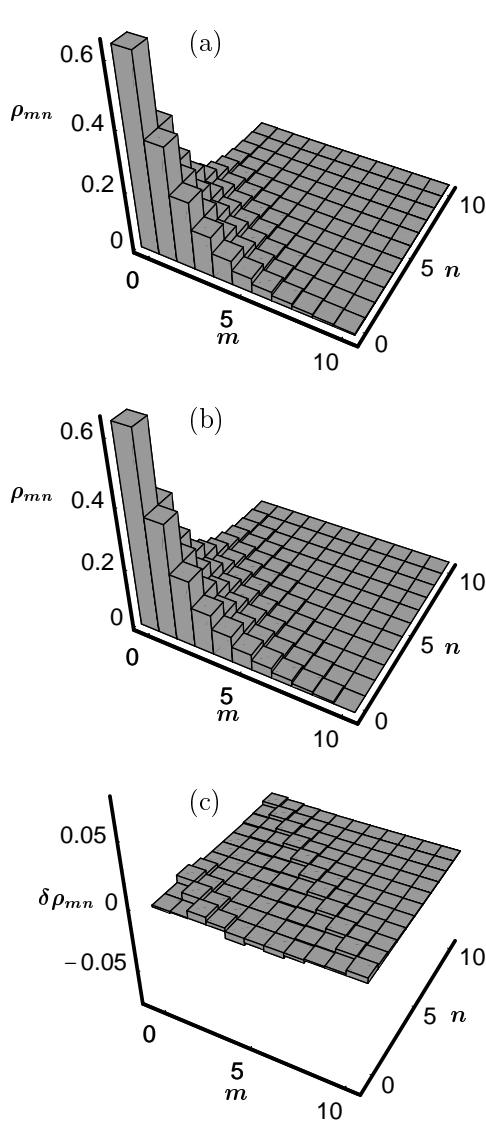}}
\caption{\narrowtext{Same as in figure 3 but for a Perelomov
coherent state
 $\Psi(0.5,0)$ (see Eqs. (17) and (18)).}}
\end{figure}
\end{center}
\vspace*{-1cm}
\subsection{Reconstruction of a coherent
superposition of pair coherent states} Our final example is the
reconstruction of a coherent superposition of pair coherent states
$\vert\Phi(\pm 3,0)\rangle$: \be
\vert\psi\rangle={e^{-i\pi/4}\over \sqrt{2}}\left[\vert\Phi
(3,0)\rangle +\vert\Phi (-3,0)\rangle\right]. \ee  It can be
easily shown that the $nm$-th density matrix element of
$\vert\psi\rangle$ will be non-zero only when both $n$ and $m$ are
even in which case its value will equal the $nm$-th density matrix
element of $\vert\Phi(\pm 3,0)\rangle$. As a result, only {\it
even} values of $k$ and {\it even} powers of $y$ appear in the
modeling of $f_k(0,y)$. As shown in Fig. 5, satisfactory agreement
is obtained between the exact and reconstructed density matrix
elements. \vspace*{-1cm}
\begin{center}
\begin{figure}[h]
\epsfxsize 7.0 cm \centerline{\epsfbox{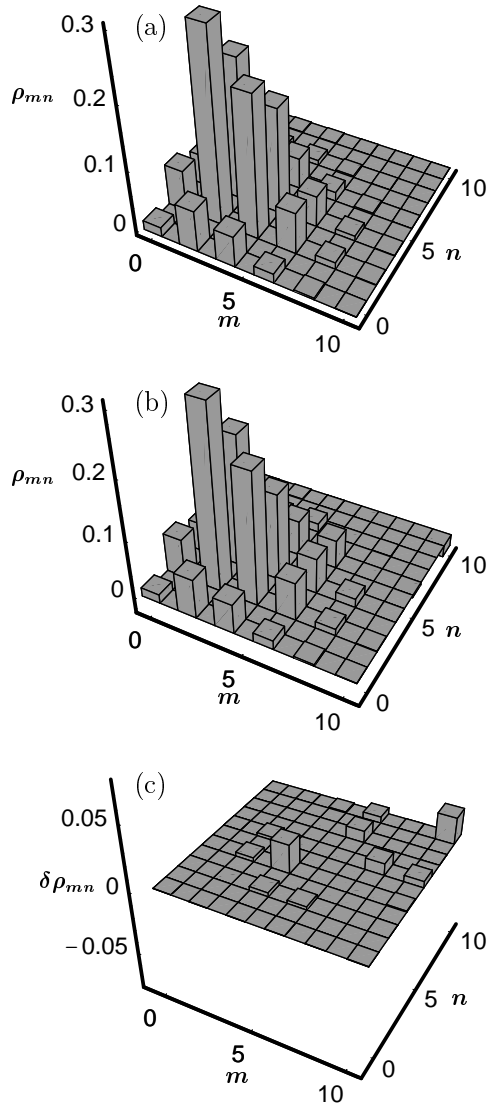}}
\caption{\narrowtext{Same as in figure 3 but for a coherent
superposition of pair coherent
 states $\vert\Phi (\pm 3,0)\rangle$ (see Eq. (19)).}}
\end{figure}
\end{center}
\vspace*{-1cm}
\section{Conclusion} We have suggested a simple
scheme for the reconstruction of two-mode SU(1,1) states using
parametric amplifiers. The probability of the output state being
in a certain two-mode number state is measured. The probability
data is then `inverted' to extract the density matrix of the input
state by taking advantage of certain symmetries in the input
state.  We have shown that this inversion can be achieved by the
least squares method after singular value decomposition of the
design matrix.

\end{multicols}
\end{document}